\documentclass[runningheads]{llncs}
\usepackage[T1]{fontenc}
\usepackage{graphicx}
\usepackage[colorlinks=true,allcolors=blue]{hyperref}
\usepackage{color}

\def\orcidID#1{\smash{\href{http://orcid.org/#1}{\protect\raisebox{-1.25pt}{\protect\includegraphics{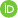}}}}}
\usepackage{booktabs}
\usepackage[%
]{todonotes}
\usepackage{csquotes}

\newcommand{\afblock}[1]{\noindent{\textbf{#1}}}

\begin{document}
\title{The (C)omprehensive (A)rchitecture (P)attern (I)ntegration method: Navigating the sea of technology}
\titlerunning{The CAPI method}
%
\author{Sebastian Copei\inst{1,3}\orcidID{0009-0009-8696-7960} \and
Oliver Hohlfeld\inst{1}\orcidID{0000-0002-7557-1081} \and
Jens Kosiol\inst{2}\orcidID{0000-0003-4733-2777}}
\authorrunning{S. Copei et al.}
\institute{Distributed Systems, University of Kassel, Kassel, Germany \\
\email{\{sco, oliver.hohlfeld\}@uni-kassel.de} \and
Philipps-Universität Marburg, Marburg, Germany
\email{kosiolje@mathematik.uni-marburg.de} \and
Innovationfield Digital Ecosystems, Fraunhofer IEE, Kassel, Germany \\
\email{sebastian.copei@iee.fraunhofer.de}
}
\maketitle              
\begin{abstract}
The technological landscape changes daily, making it nearly impossible for a single person to be aware of all trends 
or available tools that may or may not be suitable for their software project. 
This makes tool selection and architectural design decisions a complex problem, especially for large-scale 
software systems. To tackle this issue, we introduce CAPI, the Comprehensive Architecture Pattern Integration method that uses 
a diagnostic decision tree to suggest architectural patterns depending on user needs. 
By suggesting patterns instead of tools, the overall complexity for further decisions is lower as there are fewer architectural 
patterns than tools due to the abstract nature of patterns. Moreover, since tools implement patterns, 
each non-proposed pattern reduces the number of tools to choose from, reducing complexity.
We iteratively developed CAPI, evaluating its understandability and usability in small studies 
with academic participants. When satisfied with the outcome, we performed a user-study 
with industry representatives to investigate the state-of-the-art in technology selection and 
the effectiveness of our proposed method. We find that technology selection is largely performed 
via trial and error, that CAPI is uniformly perceived as helpful, and that CAPI is able 
to reproduce the productive architectural environments of our participants. 
\keywords{Architectural Patterns \and Technology Selection \and DevOps}%
\end{abstract}
\section{Introduction}\label{sec:intro}
The trend towards developing software as a (web) service promotes the creation of (large-scale) distributed systems, thereby increasing the complexity of software architectures.
Due to the complexity and scale of these systems, they are often implemented as microservice architecture and by using agile methods and DevOps organizational practices, 
such as CI pipelines, that enable teams to organize efficiently.
To facilitate the development of these systems, a wide range of software frameworks has emerged (e.g., as illustrated in the Cloud Native Landscape atlas).
These frameworks address the implementation of individual design patterns, such as API gateways, databases, and message brokers, to help simplify the developers' work in building their systems.
Despite the chosen software development methods, previous work highlights significant challenges and barriers in developing large-scale systems.
Commonly cited issues include a lack of collaboration and communication, insufficient skills and knowledge, and ineffective criticism practices~\cite{Chal1,Chal2,Chal4}.
Additionally, as we will show in this paper, developers face challenges in selecting software frameworks.
This issue primarily arises from the overwhelming number of frameworks that implement similar design patterns (e.g., message brokers) and the rapidly evolving software landscape.

In this paper, we tackle the challenge of navigating the complex landscape of software development by simplifying the process for developers to choose design patterns for large-scale software as a service projects.
Rather than recommending tools that frequently change, our approach focuses on selecting the fundamental architectural design patterns that these tools implement.
This results in a more manageable set of tools, making it easier for developers to make selections in a subsequent step.
These design patterns are foundational for building scalable software architectures, and making an informed choice is a crucial aspect of software architecture design.
Therefore, selecting the appropriate architectural design pattern is a critical step in the software architecture process.
To assist in this selection, this paper introduces the Comprehensive Architecture Pattern Integration (CAPI) method, driven by a decision tree.
To better address the needs of developers and architects, we will \emph{i)} survey their requirements and \emph{ii)} assess our CAPI method through interviews with industry practitioners.
This way, we aim to answer the following research questions:
\begin{description}
    \item[RQ1] How do practitioners select software architectures for their projects?
    \item[RQ2] What problems do the practitioners face during this process?
    \item[RQ3] How does the CAPI method help industry practitioners select better software architectural designs more easily?
\end{description}

We evaluate the CAPI method in interviews with industry practitioners and show that it addresses a need.
Our contributions are as follows.
\begin{itemize}
    \item We show that there is lack of approaches that guide the selection of software architectural designs in modern software projects.
    
    \item We contribute the CAPI method, which enables software architects to select the appropriate architectural design pattern for realizing a given software.
    The CAPI method is realized as a decision tree with questions that guide the design pattern selection process.
    The CAPI method covers the current architecture designs, with a focus on microservices as the emerging implementation approach.

    \item We evaluate the CAPI method with industry practitioners and show that it solves a technology selection problem that industry practitioners face.
    Our interviews highlight that our method is perceived as a useful approach in situations where software design patterns need to be selected.
    
\end{itemize}

%
\paragraph{Structure} 
In Section~\ref{sec:related}, we describe the related work. In Section~\ref{sec:tree}, 
we introduce the CAPI method. In Section~\ref{sec:study}, we review our qualitative study, and in Section~\ref{sec:discussion},
we discuss the results. Finally, Section~\ref{sec:conc} summarizes our work and briefly describes our next steps.
\section{Related Work}\label{sec:related}
%
%
\afblock{Architectural design selection.}
Regarding architecture selection, there is a range of studies that investigate the relation between architectural patterns and quality attributes of the resulting system~\cite{KassabEM11,MePL17,HaouesSBC17,KassabMLS18,MeijerTA24}; Haoues et al.~\cite{HaouesSBC17} even 
derive a decision model where a foundational architectural design decision is suggested depending on quality 
attributes one wants to focus on. However, their systems addresses the foundational style of architecture to be 
used in a system (model-driven architecture, component-based architecture, service-oriented architecture, 
event-driven architecture, aspect-oriented architecture), whereas we are interested in helping users to decide 
which of the plethora of patterns to use that have been suggested in connection with modern, often cloud-based 
development. In particular for component-based architectures, approaches exist to automatically optimize an architectural 
design, given as a Palladio component model~\cite{KoziolekKR11,BuschFK19}. 
Razzaq extracts patterns for microservice development from the literature and maps these to common problems 
in IoT development they can help to solve~\cite{Razzaq20}. 
Decision support systems (DSS) for technology selection have also been developed in other contexts, e.g., for database technologies~\cite{DSS1}.
Finally, in the paper most similar to ours, Farshidi and Jansen develop a DSS that supports
the decision-making process for pattern-driven software architecture~\cite{DSS2}.
Whereas our method relies on a decision tree fueled by diagnostic questions, the DSS mentioned in the 
publication uses a list of definable requirements prioritized by the Must have, Should have, Could have, 
and Won't have (MoSCoW) technique to determine fitting patterns. The authors use an inference engine to generate
a list of patterns that fulfill the requirements. Similar to our approach, the authors extract the patterns used
in their method from a literature review. In total, the authors used 29 patterns with 40 features. From this set,
the inference engine can list fitting patterns depending on the features that will fulfill a requirement. 
Our method uses a set of 47 patterns and creates a result set due to a 
decision tree instead of an inference engine.  In comparison, the DSS and our method share four patterns,
meaning there are 25 patterns we are not including and 44 they do not include. 
We explain these deviations with the methodology of the literature reviews performed beforehand.
Nevertheless, from this deviation, we are confident that our method can cover a broader set of target architectures 
as the number of patterns it can suggest is significantly higher. Moreover, the extendability of an inference engine, 
if they want to add more patterns, is questionable as the complexity rises for each pattern added.
In contrast, the change or the extension of our decision tree is easily possible by adding further 
questions or even extra paths.

We observe that none of the surveyed works address the problem of architectural decision-making for current software stacks in software as a service projects.
Other methods do not build on the same pool of patterns as ours and address different levels in the design space.
We aim at closing this gap with the CAPI method.
\section{The CAPI method}\label{sec:tree}
In this section, we describe the decision tree (Sect.~\ref{sec:final_tree}) that constitutes the heart of the CAPI method 
and present its iterative development process (Sect.~\ref{sec:tree_dev}).
\subsection{CAPI's decision tree}\label{sec:final_tree}
CAPI's overall approach involves asking users a series of questions, organized in a decision tree format.
Based on their answers, it provides a set of architectural design patterns to implement the envisioned system.

\begin{figure}[th]
    \centering
    \includegraphics[width=.65\textwidth]{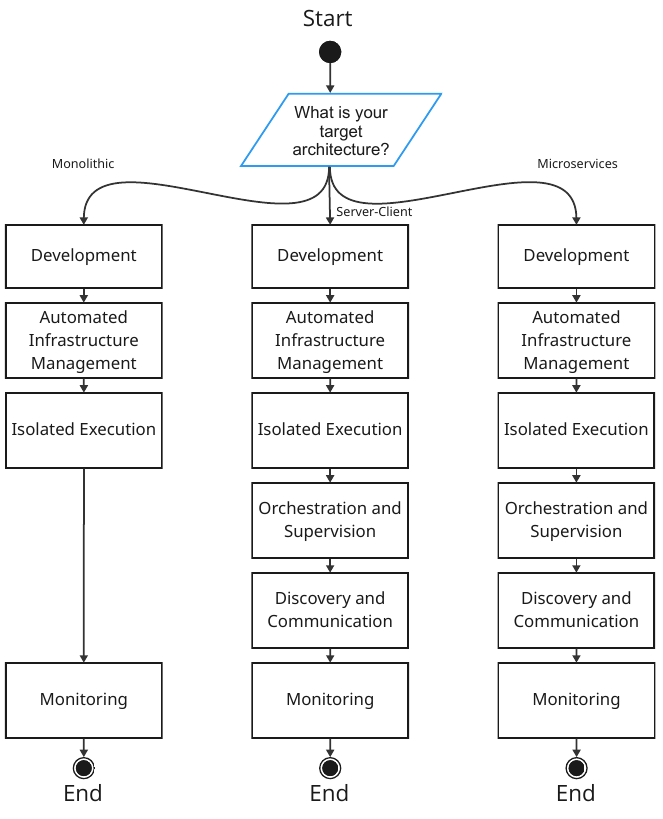}
    \caption{Abstract structure of CAPI's decision tree using categories}
    \label{fig:tree_cat}
\end{figure}

\afblock{Assumptions and entry point of CAPI.}
\emph{In our approach, we assume that a decision for a \emph{foundational architectural style} has already 
been made.} That is, it has been decided to implement the software as monolithic, client-server or microservices.
The process of making this decision is beyond the scope of this paper.
Depending on the selected foundational architectural style, a user is guided through a 
series of questions about the envisioned system; the given answers determine a set of \emph{architectural 
design patterns} that are suitable to implement the envisioned system in the selected foundational architectural style. 
The overall abstract structure of the decision tree is depicted in Fig.~\ref{fig:tree_cat}. 

The initial question of our decision tree asks for the foundational architectural style, with 
\enquote{monolithic}, \enquote{client-server} and \enquote{microservices} being possible answers. 
We opted to not support users regarding that fundamental decision for two reasons:
First, as seen in Section~\ref{sec:related}, such fundamental decisions have at least partially been addressed 
in prior work. And, more importantly, in our experience these fundamental decisions are not only influenced by 
technical considerations but also heavily by cultural (e.g., knowledge and openness of development team) or economical 
(e.g., \enquote{Which resource can we afford to use for this system?}) ones. Thus, the foundational decision
deserves its own, multi-faceted process that can be distinguished from a process that helps with an appropriate 
technical implementation, on which we want to focus in this work. 
We support \enquote{monolithic}, \enquote{client-server} and \enquote{microservices} as foundational 
architectural styles because, in our experience (and also suggested by the results of Kassab et al. in~\cite{KassabMLS18}), 
these are widespread and considerably different ways in which to develop a modern service-oriented system. 

\afblock{Structure of CAPI's decision tree.}
Generally, the flow of the decision tree is structured by six \emph{categories}, into which 
architectural design patterns can be grouped (compare again Fig.~\ref{fig:tree_cat}). 
Iteratively, for each category and for each pattern that belongs to it, a series of 
questions is asked to determine whether or not the respective pattern should belong to the 
result set. Currently, the tree consists of three paths: Each possible answer to the initial 
question about the fundamental architectural style leads to a distinct path in the tree 
because in each case, there is a different set of applicable patterns to ask for.

\begin{figure}[th]
    \centering
    \includegraphics[width=\textwidth]{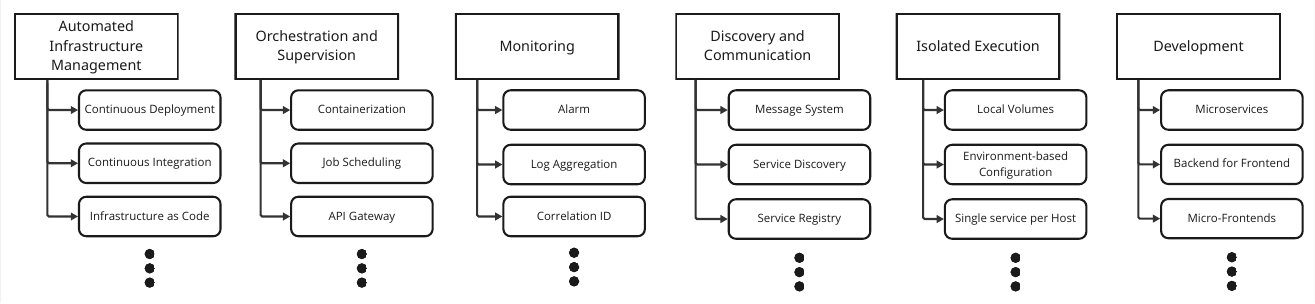}
    \caption{Categories and Patterns used to create the decision tree}
    \label{fig:cat_pat}    
\end{figure}

Our selection of design patterns is based on those identified from our recent survey study~\cite{RapidReview}, 
where we systematically collected architectural design patterns for DevOps and microservices.
From the 52 reported patterns, we use 47 to design the decision tree and group these into 
six different categories, namely \emph{Development}, \emph{Automated Infrastructure Management}, 
\emph{Isolated Execution}, \emph{Orchestration and Supervision}, \emph{Discovery and Communication} 
and \emph{Monitoring}. Our grouping of patterns is inspired by the well-known DevOps cycle. 
An overview of the categories and examples for patterns belonging to these is provided in 
Fig.~\ref{fig:cat_pat}. 

For each pattern of each category that is relevant
for the selected architectural style, a series of two to three questions decides whether the decision tree 
suggests to use that pattern; sometimes such a series of questions also checks for a small group of
patterns that should be used together. 
The first question(s) always check whether there are positive reasons to use 
the pattern(s) to develop the envisioned system. The last question in such a series is a \emph{contraindication} 
question to ensure that there is no reason to not use the pattern(s). 
For simplicity, the questions can only be answered with \enquote{yes} or \enquote{no}. If a user's answer 
to the first question(s) already indicates that a pattern is not applicable, the remaining question(s) 
for that pattern are skipped and the process immediately continues with the questions for the next pattern.  
To not influence a user in one way or other, they are only presented the questions, without being 
explicitly told for which pattern the current question checks. The relevant patterns are collected in 
the background and only presented to the user as result when they traversed the whole tree. 


\paragraph{\textbf{Illustration of questions and usage.}}
We use the following scenario to illustrate a snippet of the tree's usage. 

{\footnotesize
\emph{Assume you work in a large software engineering team at \emph{Streamland}, a newly founded company that 
has acquired most streaming licenses for movies and series worldwide. Streamland wants to build a new platform that 
combines all the current streaming services so customers only need a single service to stream all movies and series. 
The application should be available for all devices, like TVs, smartphones, or PCs.}
}

\begin{figure}
    \centering
    \includegraphics[width=\textwidth]{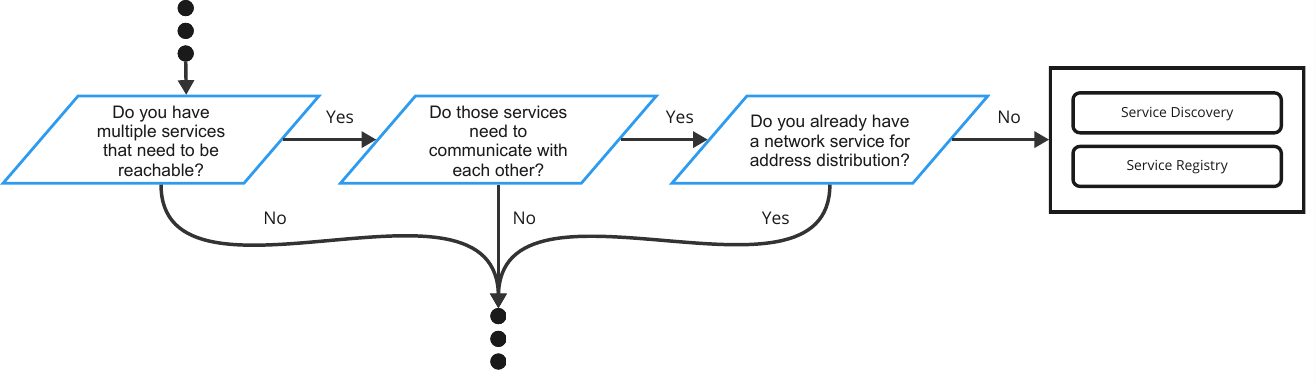}
    \caption{Example question flow}
    \label{fig:tree_example}
\end{figure}
As shown in Fig.~\ref{fig:tree_cat}, the first decision is to choose the overall architectural style. 
In this example, we choose a microservice architecture. Given that decision, the user is asked questions 
for patterns that belong to the \enquote{Discovery and Communication} category (which is not the case for 
a monolithic architecture), amongst which there are \enquote{Service Discovery} and \enquote{Service Registry}.
Figure~\ref{fig:tree_example} shows the question flow for these two patterns. 
Here, the first two questions ask whether the patterns are needed, and the third question is the contraindication question.
Only answering the first two questions with \enquote{yes} and the third one with \enquote{no} will lead to the result that the two patterns are added to the result set of suggested patterns.
Otherwise, we will continue to the next question flow. This concept is applied to all question flows for each path in the tree. 
In the case of our example, as we want to implement a microservice-based architecture we need to use multiple services that need 
to communicate with each other. If we already have infrastructure that hosts mechanics for service discovery, like Kubernetes, 
the decision tree will send us to the next question flow. Otherwise, it would suggest the patterns for \emph{Service Discovery} and 
\emph{Service Registry}. CAPI's full decision tree is available in our research bundle~\cite{zenodo}.

\afblock{Implementation.} 
To evaluate the CAPI method, we developed a simple online tool, in which a user can 
click through the questions. In the end, the tool displays the suggested patterns. 
The tool is available online~\cite{zenodo}. 

\afblock{Discussion.}
The CAPI method is intentionally focusing on selecting the architectural design patterns that are relevant in realizing the intended software.
This way, we intentionally do not focus on selecting the foundational architectural style (e.g., monolithic vs.\ microservice), since this decision is bound to other factors (e.g., current practices in the software team, etc.) and already addressed by prior work.
Furthermore, we intentionally focus the decision tree on the current three dominant architectural styles (i.e., monolithic, client-server, and microservice).
The approach described in this work can be used to extend the decision tree in future work to cover a broader set of styles.
Extending the tree can be done by adding more branches.
In that, it is even possible to gain (large parts of) the new branch 
by taking the most comprehensive branch of our tree, the one for microservices, and dropping all 
questions that check for patterns irrelevant to the considered style. 
In fact, we partly developed our tree in that way, highlighting that extensions are easily possible. 
\subsection{Iterative development process and first quantitative analysis}\label{sec:tree_dev}
We developed our decision tree in two iterations and concluded each of the two phases with a small-scale quantitative study 
to determine the intelligibility of our questions and to get a first impression of the quality of architectural designs 
users obtained with it. 
The two studies were performed with disjoint groups of participants recruited from academia. 
In each case, the participants should apply the tree in its form at that time in four artificial scenarios of different 
complexities (including, e.g., the Streamland scenario from above as most complex system). 
For both these studies, the participants were provided with the questions of the tree printed on paper. 
In the end, they were presented the sets of patterns that resulted from their answers for each scenario and then asked to answer 
a questionnaire. There we asked questions (some open, some answerable on a Likert-scale) to determine 
how the participants used the tree, as how intelligible the tree's questions and as how helpful its usage was perceived 
to arrive at a design, and we asked for general feedback. Moreover, we compared the resulting architectures that the participants 
obtained to architectures that we perceived as appropriate for the scenarios.

In our initial draft, we already used the overall structure of the tree as described above 
and the same set of patterns. However, per pattern, we asked a single question that checked 
for existence of a reason to use that pattern. 
We performed our first study with a group of seven academics, consisting of 
PhD students, PostDocs and professors. The participants perceived the tree as helpful and 
rated the intelligibility of the asked questions as high. 
However, when we compared the resulting designs to our intended ones, we observed that 
the participants tended to answer with \enquote{yes}, leading to unnecessary large sets 
of patterns for many scenarios. 

We explained that phenomenon with the first draft of the tree not associating any kind of 
\enquote{cost} with selecting a pattern. 
Besides minor rephrasing of individual questions, we tackled that issue with (i) asking for 
several reasons to use a pattern in the case of complex ones and (ii) introducing our 
contraindication questions that check for the existence of a reason to not use a pattern. 
We performed a second user study with exactly the same setup as the first, now using the revised
tree. We also recruited our second set of participants (ten) from academia, now consisting of 
advanced undergraduates and master and PhD students. 
Despite being far less experienced (academically and in terms of DevOps and cloud-based 
development), the phenomenon of selecting a large amount of unnecessary patterns was 
noticeably reduced. The decision tree was still largely perceived as intelligibly and 
helpful. Therefore, we just used the additional feedback of that study to clarify a few 
further questions and used the resulting design to evaluate our approach with 
industrial practitioners, as presented next.
\section{CAPI in action: An industrial user study}\label{sec:study}
We are now interested in examining the satisfaction levels of industry professionals with CAPI.
To do so, we have conducted interviews with industrial representatives to evaluate whether CAPI can help select better software architectures.
We are further interested to understand how technology selection is performed in industrial practice, 
which challenges practitioners face, and if the CAPI method helps in dealing with these challenges (RQs~1--3).

\subsection{Methodology}\label{sec:method_study}
We base our study on using a mix of structured and semi-structured interviews.
We used predefined questions in a fixed order but allowed participants to answer freely. 
The interviews ended with a case study in which the participants had to use CAPI via a web interface.
All interviews were performed by the first author (individually with each participant) and recorded with Microsoft Teams. 
Microsoft Teams also created the initial transcript of the interviews. Afterward, we refined the transcripts by hand to 
improve their readability. During the data collection phase, we extracted the core statements from the participants' 
answers using the refined transcripts and compiled them into a spreadsheet. The transcript of the interviews 
and our data collection files are available at Zenodo~\cite{zenodo}.

\afblock{Participant selection.}
As DevOps is an emerging technology for developing large-scale applications, 
we focused on participants who use DevOps for their application development. Furthermore, we separated the participants into two groups. 
In the first group, participants had not adopted DevOps but concretely planned to do so for a near-term project (Group~1 (\textbf{G1})). 
Participants from the second group are actively using DevOps for their projects (Group~2 (\textbf{G2})).
By separating the participants into these two groups, we aim to better understand the impact of decisions
made in the initial phase of projects. These insights could affect the positioning of the CAPI method from a 
timing perspective. For example, is it enough to only use the method once in the beginning, 
or may it be better to recheck the decisions with the method in an iterative manner?
We recruited the participants from the first author's network and country and ensured that the resulting participants 
were distributed among multiple companies and different branches of business.

\afblock{Study procedure.}
The procedure of the interviews consisted of the following steps: 
\emph{i)} Initial questions about the technology choice process and potential problems; 
\emph{ii)} Participants need to determine the technologies they need using the CAPI method; 
\textbf{G1} does this for the project they want to realize, and \textbf{G2} for the project they already have implemented; and 
\emph{iii)} Questions about the usage of the CAPI method.
We asked participants from \textbf{G1} the following questions before they used our method (\textbf{Q1}, \textbf{Q2}) and after they used our method (\textbf{Q3}, \textbf{Q4}).
\textbf{Q1}: Do you have a process for identifying necessary technologies?
\textbf{Q2}: Could there be problems with this process?
\textbf{Q3}: Would the tree have helped you with your problems in identifying necessary technologies?
\textbf{Q4}: Would the result of the tree now help you implement DevOps?
We also asked participants from \textbf{G2} about their technology choice process and their experiences with problems. 
Additionally, we asked how strongly the results of our method differ from their actual architecture. The questions are as follows. 
\textbf{Q5}: How did you select the necessary technologies?
\textbf{Q6}: Have you encountered any problems with this approach?
\textbf{Q7}: Would the tree have helped you with your problems in identifying necessary technologies?
\textbf{Q8}: How much does the result of the tree differ from the stack you are actually using?
\subsection{Results}\label{sec:result}
We interviewed ten participants from six companies in Germany, which consist of software development service providers and consultant firms.
We interviewed participants from the same company regarding different projects. The participants are evenly distributed across the two groups. 
Five participants plan to use DevOps, and five are actively using it. 
Table~\ref{tab:part} overviews our interview partners' demographics and assignment to study groups.
\begin{table}
    \centering
    \caption{Overview of survey participants}
    \label{tab:part}
    \begin{tabular}{@{}ccllr@{}}
        \toprule
        Participant ID & Company ID & Branch & Position & Group \\
        \midrule
        P1  & C1 & Heat Industry    & Software Engineer        & 2 \\
        P2  & C1 & Heat Industry    & Software Engineer        & 2 \\
        P3  & C2 & Energy Economics & Software Engineer        & 2 \\
        P4  & C3 & Energy Economics & Software Engineer        & 1 \\
        P5  & C4 & SAP              & Technical Consultant     & 1 \\
        P6  & C4 & SAP              & Technical Consultant     & 1 \\
        P7  & C5 & Multiple         & Cloud Solution Architect & 2 \\
        P8  & C6 & Consultant       & Lead Architect           & 1 \\
        P9  & C5 & Multiple         & Cloud Engineer           & 2 \\
        P10 & C5 & Multiple         & Lead DevOps Engineer     & 1 \\
        \bottomrule
    \end{tabular}
\end{table}
Table~\ref{tab:intresbd} summarizes the results of the interviews with the participants from G1 (about to adopt), 
and Table~\ref{tab:intresad} the ones for G2 (already using). 

\afblock{Lack of a architecture selection process.}
Both the adopters and non-adopters stated they do not have a process to identify needed technology (Q1, Q5).
In detail, G1 reported that the technology was specified by their clients (P6, P10) or that they relied on experience
and iterated their decision using this experience (P4, P5, P8, P10). G2 also reported that their clients specified 
technology (P1, P2, P7, P8). Further, they named trial-and-error (P3, P9) and self-motivated employees (P9) 
as drivers for their technological decisions.
This highlights the need for a guided technology selection approach such as CAPI.

\begin{table}
    \centering
		\footnotesize
    \caption{Interview results for Group~1 (about to adopt)}
    \label{tab:intresbd}
    \resizebox{\textwidth}{!}{%
    \begin{tabular}{@{}llr@{}}
        \toprule
        Question & Answers & Participant \\
        \midrule
        Q1	& No                                                   		                           & P4, P5, P6, P8, P10 \\
            & \quad Technologies were specified by client                                            & P6, P10 \\
			& \quad We rely on our experience and do it iteratively                                  & P4, P5, P8, P10 \\
        \midrule
        Q2	& No, we do not encounter any problems                    	                           & P4 \\
            & By using the already known, we miss new things                                        & P5, P10 \\
            & As the knowledge is widely distributed, it is cost-intensive to collect it                  & P10 \\
        \midrule
        Q3	& Yes                                                                                  & P4, P5, P6, P8, P10 \\
            & \quad It hints in the correct direction                                              & P4, P6, P8 \\
			& \quad But as our projects are inflexible, the results need to be complete            & P5 \\
            & \quad But the method should also suggest tools for the patterns                      & P6, P10 \\
            & \quad But the method should show the reasoning for a pattern                         & P10 \\
        \midrule        
        Q4	& Yes, the method would have helped us                                                 & P5, P6, P8, P10 \\
            & \quad The method is a good checklist                                                 & P5, P8 \\
            & \quad It would be even better if the method showed the relationship between patterns & P8, P10 \\
            & \quad The method should be used multiple times during an iterative process           & P8 \\
            & No, as we do not know all of the patterns                                            & P4 \\
        \bottomrule
    \end{tabular}}
\end{table}

\afblock{CAPI is not for everyone.}
Not every industry practitioner \emph{directly} faced technology selection problems, e.g., since the technology selection was made by the client.
Regarding the problems encountered during the technology selection process (Q2, Q6), 
G1 reported that they did not encounter any problems (P4), they missed out on new things (P5, P10), 
and high costs for knowledge collection (P10). G2 reports similar experiences; 
they also did not encounter any problems as their clients specified everything so there was no choice (P1, P2), 
high costs were a problem (P3, P9), and finding the correct point in time to move to the next technology if
the currently selected one proves to be unsuitable.
This highlights that not every industry practitioner needs to use CAPI, in particular when the technology selection is provided by clients.
In these cases, however, the problem is simply shifted since the clients need to select the technology.

\afblock{Does CAPI solve a problem?}
After using the CAPI method and asking if it could help them solve their problems (Q3, Q7), 
all participants from G1 and G2 agreed that it could and would help them during their technology selection process.
Pointing to the right direction (P4, P6, P8) and helping to focus on the actual requirements (P3, P7) 
are the highlighted positive aspects of the CAPI method.
Additionally, participants from G1 also reported that the method should be nearly complete as their project is very
inflexible (P5). After suggesting patterns, the method should suggest tools that implement the patterns (P6, P10),
and it should show the reasoning for a selected pattern (P10).
G2 reported additionally that the method should suggest tools for the patterns (P1, P2, P3, P7), 
the method should be used multiple times in an iterative process (P3), and it would be helpful if, besides the patterns, 
the relationship between them should also be highlighted (P9).
\begin{table}[t]
    \centering
    \caption{Interview results for Group~2 (adopters)}
    \label{tab:intresad}
    \resizebox{\textwidth}{!}{%
    \begin{tabular}{@{}llr@{}}
        \toprule
        Question & Answers & Participant \\
        \midrule
        Q5	& There was no process, as our client specified many factors                                     & P1, P2, P7, P8 \\
            & We just tried out the technologies that were not specified                                     & P3, P9 \\
            & The knowledge comes from self-motivated employees                                              & P9 \\
        \midrule
        Q6	& No, as everything was specified    					                                         & P1, P2 \\
            & The process of trying out was very cost-intensive                                              & P3, P9 \\
            & It is hard to find the correct timing to try the next technology if there are problems         & P7 \\
        \midrule
        Q7	& Yes                                                                                            & P1, P2, P3, P7, P9 \\
            & \quad If the method also would suggest tools for the patterns                                  & P1, P2, P3, P7 \\
            & \quad It helps focus on what is needed     											         & P3, P7 \\
            & \quad But only for greenfield projects without client specifications                           & P9 \\
            & \quad The method should be used multiple times during an iterative process          		     & P3 \\
            & \quad prioritization and viewing of the relationship between the patterns would help even more & P9 \\
        \midrule
        Q8	& On the first view, the results match with our stack          		                             & P1, P2, P3, P7, P9 \\	
            & Some patterns that we are using are not highlighted 									         & P7, P9 \\
            & Methods highlight even patterns that we plan to implement								         & P3 \\
        \bottomrule
    \end{tabular}}
\end{table}

When asked if the method would be helpful for their current project (Q4), 
four participants from G1 answered \enquote{yes} (P5, P6, P8, 10), and one from G1 answered \enquote{no} (P4). 
The participant who answered \enquote{no} explained that he did not know all the patterns listed by the method.  
Furthermore, the participants from G1 stated that the method is a good checklist (P5, P8), 
the relationship between patterns should be highlighted (P8, P10), and that the method should be used multiple
times during an iterative process (P8). 

\afblock{CAPI provides a sound technology selection.}
Finally, we asked the participants from G2 about the differences between the patterns selected using the 
CAPI method and the actual patterns used in their project (Q8). All five participants agreed that the patterns 
suggested by our CAPI method match the actually used patterns (P1, P2, P3, P7, P9). 
Complementary, two participants (P7, P9) stated that a few patterns were not selected. 
P9 concretized that the pattern \emph{Infrastructure as Service} was not suggested by the CAPI method. 
For one project, our method suggested three more patterns (\emph{Fault Injection}, \emph{System-wide Resiliency}, \emph{Log Aggregation})
that were not implemented yet but should be in the future (P3). 
\section{Discussion}\label{sec:discussion}
In this section, we interpret the obtained results to answer our research questions.
\subsection{Teams have no dedicated structural process to select technology for their projects (RQ1)}
All ten participants from the two groups stated they do not have a dedicated, 
structured process for selecting technologies for their projects. However, two participants of G1 and four of G2 explained 
that they cannot select technology (completely) independently, as their clients specify (at least a partial) technology stack. 
We interpret this finding as the participants not needing to select (all) technology. Thus, they have no process. 
On the other hand, four members of G1 and three of G2 also stated that they relied on experienced employees (P4, P5, P8, P9, 10) 
or just tried technology until they found one that was suitable for a specified requirement (P3, P9). 
These answers indicate that if there are no external regulations, like a client's specifications, 
the actual process of technology selection relies on experienced employees.
Although six of ten participants stated that they did not have a process for technology selection as their clients specified the technology, 
only three participants (P1, P2, P4) reported that they did not encounter any problems under this circumstance. 
As the clients do not fully specify the whole technology stack and often work with allow-lists, 
the companies can still choose a certain amount of technology, which again leads to problems. 
Even if the whole stack is specified, the actual problem of technology selection then shifts from the implementing company to the specifying client.

All in all, for RQ1, we extracted from our interview that there is often no process for selecting 
technology because the clients specified these technologies beforehand. Regardless of whether the client regulates the technology to use, 
the companies rely on their experience and reuse technology already known or use trial-and-error to determine the well-suited technology.
\subsection{Selecting technology without a process can lead to time- and money-consuming trial-and-error (RQ2)}\label{sec:req2}
Both groups reported that a trial-and-error-based process is very costly (P2, P9, P10). Furthermore, 
these costs may increase if the knowledge is spread over multiple persons as the knowledge needs to be collected (P10). Moreover, 
G1 reported that they tend to reuse already known technology instead of trying new ones, 
leading to missing out on new technology and trends (P5, P10). As Participant P10 reports, on the one hand, 
they miss out on new technology because they stick to the already known, and on the other hand, 
the process of knowledge collection is cost-intense. In our opinion, these two problems could be interconnected. 
As technology collection and discovery of new ones is a cost-intensive investment, 
this problem could automatically lead to sticking to the technology that the company's employees already know because, 
through this behavior, there is no need to invest money into a technology selection process. Additionally, 
G2 reports that if there is a new technology, it will only be introduced by employees with a certain self-motivation to 
encourage a specific technology (P9). We interpret this also as a symptom of the high costs of a technology selection process, 
so companies instead rely on self-motivated employees to reduce costs.

In conclusion, the answer to RQ2 is that if companies have no dedicated structured process for selecting needed technology,
they tend to do trial-and-error or, as this is very costly, rely on self-motivated employees who bring this knowledge without extra costs. 
However, in both ways, either the companies have to pay for a costly trial-and-error process, or they need to allocate valuable, 
knowledgeable employees to the process, leading to a higher workload for these specific people. 
\subsection{CAPI, the map for the technology labyrinth (RQ3)}
We organize this research question in multiple blocks of key takeaways and summarize in the final block our conclusion for RQ3.

\afblock{From a first impression, the method supports practitioners during the technology-choosing process.}
Nine of the ten participants concluded that the CAPI method would be helpful as it closes the gap of the missing process of
selecting technology for a project. Only participant P4 declared that the CAPI method would not help him as neither he nor his team were fully aware of all patterns 
that were suggested by the method. As reported in Section~\ref{sec:req2}, the main issue with a missing process is that the needed
technology is selected through a trial-and-error process, which leads to increased costs. Although the participants do not directly address this issue,
they reported that the method is a good checklist (P4, P8), helps focus on the things that are needed (P3, P7), 
and helps by hinting the focus in the correct direction (P4, P5, P8). 
These findings imply that the CAPI method provides direction in the technology landscape. 
With a navigation tool, the technology-choosing process is less time-consuming and more structured than the previously mentioned trial-and-error process,
leading implicitly to reduced cost.
We conclude that the CAPI method hit the right nerve through our approach of suggesting 
patterns instead of tools and is indeed helpful for the technology selection process.
\afblock{Software Engineers prefer to get tool suggestions in addition to the patterns.}
From G1, the reported potential improvements are that the method should also suggest tools that implement the patterns (P6, P10), 
the method should show the reasoning for a selected or not-selected pattern (P10), the method should also highlight the relationship between selected patterns (P8, P10), 
and the method should be rather used multiple times during an agile development process instead of once in the beginning (P10). 
Also, from G2, participants report that the capability of suggesting tools that implement the selected patterns would be a desirable improvement (P1, P2, P3, P7).
Furthermore, like G1, highlighting the relationship between patterns (P9) and the usage of the method multiple times during an agile process (P3), 
G2 saw these improvements, too. Additionally to G1, G2 also mentioned that in addition to highlighting the relationship of the patterns, 
a prioritization of the patterns would be helpful to determine which patterns should be implemented first (P9).
Surprisingly, six out of ten participants mentioned that the method should also suggest tools in addition to patterns. 
This finding seems to conflict with our thought that the strength of our approach was to suggest patterns instead of tools, 
as the landscape of tools is far broader than the available patterns. However, P1, P2, and P3 declared themselves as Software Engineers who 
develop software for their projects, so for them, the choice of tools is more important as they will directly use them. 
The other participants working on a more abstract layer have not mentioned the necessity of suggested tools besides P7. 
From this perspective, the result does not conflict with our approach. 
We conclude that the desired improvements for the CAPI methods are: 
i) for selected and non-selected patterns there should be a reasoning, 
ii) selected patterns should be prioritized, and their relationship should be visualized, 
iii) the method should be used multiple times during the development instead of just in the beginning, 
iv) additionally to the selected patterns, tools should be suggested that implement the patterns.
\afblock{Regardless of the role in the project, the CAPI method is able to reproduce a productive environment.}
All G2 participants reported that the set of patterns resulting from applying the CAPI method roughly 
matches their actual implemented stack (P1, P2, P3, P7, P9). 
Additionally, one participant mentioned that the resulting set of suggested patterns contains patterns they have not used but want to implement in the future (P3). 
However, the CAPI method could not reproduce the whole stack of two participants because it missed some patterns (P7, P9).
Nevertheless, these findings make us confident that our approach is a valuable solution to fill the gap if there is no 
dedicated process to select the needed technology in a project. Interestingly, both Software Engineers (P1, P2, P3) and
Architects (P7, P9) stated that the results nearly match from their point of view. We interpret these as,
whether a more tool-driven software engineer or an abstract architect, both groups were satisfied with our method on the baseline.
From our interviews, we derive that the expected output of such a technology choice method is tools for software engineers. 
Nevertheless, they do not reject the approach of suggesting patterns; instead, they are able to find their used tools in the patterns.
\afblock{Finally, we answer RQ3 as follows.}
The reported problems in our interviews are high costs due to an unorganized trial-and-error process for determining suitable technology for their requirements. 
Our method is able to reproduce the participants' production environments, so the suggested patterns are trustworthy. 
Therefore, the method could lower the initially mentioned cost due to time savings, as the method suggests patterns that are trustworthy and implicitly press the user in a specific direction,
which reduces the number of tools to be investigated. By following the concerns of the participants and suggesting tools that implement the patterns additionally, 
the cost reduction may be even higher. Equipped with the CAPI method as a map, both architects and software engineers 
can benefit from the structure it offers. As mentioned by the participants, this is not fully limited to greenfield projects; 
the method can be used multiple times in an agile process, making it also suitable for brownfield projects.
\subsection{Threats to validity}\label{sec:val}
The main threat to the \emph{external validity} of our study is the small sample size and the recruiting method for our subjects. 
We cannot be sure about the completeness of received answers (e.g., on reasons for the (non-)helpfulness of our method) and how well our results generalize to other settings. 
Still, we recruited participants in considerably different branches and roles and, for some questions, received very homogenous answers. 
Therefore, we are confident that our most important findings (regular lack of technology selection process and perceived helpfulness of the CAPI method) will also hold in other settings (although possibly to a different extent).
The main threat to \emph{internal validity} is that the first author individually performed the interviews and data extraction, 
increasing the threat of bias and misinterpretation. 
To mitigate this threat, all authors together discussed examples for data extraction and the whole interpretation of the 
obtained data.
Moreover, for lack of a closely comparable alternative process, our industrial study focused on the usage of the CAPI method. 
In this way, we cannot (and do not) claim superiority over other methods or draw conclusions about which aspects of our method affect its perceived helpfulness and how. 
In any case, the widespread lack of methodological approaches to technology selection in industry highlights the importance of developing such.
\section{Conclusion}\label{sec:conc}
This paper introduces the CAPI method, which suggests architectural patterns using a decision tree with diagnostic questions about the user's different requirements.
We conducted qualitative interviews to investigate the need for such a method and its effectiveness. From these interviews, 
we conclude that the method hits the correct spot and is a possible solution for a missing process for selecting technology in a project.
Besides the positive feedback from our interview participants, we also got valuable improvements that could increase the effectiveness of our method. 
From these improvements, we derive our next steps, which consist of investigating relationships between our patterns, 
including a possible prioritization derived from these relationships, and developing a method to map tools onto patterns efficiently. 
In addition, a revaluation of the improved method would provide us with meaningful insight to determine the effectiveness of the changes.
%
%
%
\begin{credits}
\subsubsection{Data Availability.}
Our research bundle is available at Zenodo~\cite{zenodo}. 
The online tool used during the interviews is available at \url{https://capi.uniks.de}.
The tool's authentication is part of the research bundle, follow the readme for instructions.

\subsubsection{\discintname}
The authors have no competing interests to declare that are relevant to the content of this article.
\end{credits}
%
%
%
%
\bibliographystyle{splncs04}
\bibliography{anon}

\begin{thebibliography}{10}
\providecommand{\url}[1]{\texttt{#1}}
\providecommand{\urlprefix}{URL }
\providecommand{\doi}[1]{https://doi.org/#1}

\bibitem{BuschFK19}
Busch, A., Fuchss, D., Koziolek, A.: {PerOpteryx: Automated Improvement of Software Architectures}. In: {IEEE} International Conference on Software Architecture Companion. pp. 162--165 (2019)

\bibitem{RapidReview}
Copei, S., Kosiol, J.: {DevOps Patterns: A Rapid Review}. In: Software Architecture. ECSA 2023 Tracks, Workshops, and Doctoral Symposium. pp. 33--50 (2024)

\bibitem{zenodo}
Copei, S., Kosiol, J., Hohlfeld, O.: {Research bundle for "The (C)omprehensive (A)rchitecture (P)attern (I)ntegration method: Navigating the sea of technology" } (Mar 2025), \url{https://doi.org/10.5281/zenodo.15105538}

\bibitem{DSS2}
Farshidi, S., Jansen, S.: {A Decision Support System for Pattern-Driven Software Architecture}. In: Software Architecture. pp. 68--81 (2020)

\bibitem{HaouesSBC17}
Haoues, M., Sellami, A., Ben{-}Abdallah, H., Cheikhi, L.: {A guideline for software architecture selection based on {ISO} 25010 quality related characteristics}. Int. J. Syst. Assur. Eng. Manag.  \textbf{8}(2s),  886--909 (2017)

\bibitem{Chal2}
Kalantari, R., Lethbridge, T.C.: {Unveiling Developers' Mindset Barriers to Software Modeling Adoption}. In: 2023 ACM/IEEE International Conference on Model Driven Engineering Languages and Systems Companion. pp. 737--746 (2023)

\bibitem{KassabEM11}
Kassab, M., El{-}Boussaidi, G., Mili, H.: {A Quantitative Evaluation of the Impact of Architectural Patterns on Quality Requirements}. In: Software Engineering Research, Management and Applications. vol.~377, pp. 173--184. Springer (2011)

\bibitem{KassabMLS18}
Kassab, M., Mazzara, M., Lee, J., Succi, G.: {Software architectural patterns in practice: an empirical study}. Innov. Syst. Softw. Eng.  \textbf{14}(4),  263--271 (2018)

\bibitem{Chal1}
Khan, M.S., Khan, A.W., Khan, F., Khan, M.A., Whangbo, T.K.: {Critical Challenges to Adopt DevOps Culture in Software Organizations: A Systematic Review}. IEEE Access  \textbf{10},  14339--14349 (2022)

\bibitem{Chal4}
Khattak, K.N., Qayyum, F., Naqvi, S.S.A., Mehmood, A., Kim, J.: {A Systematic Framework for Addressing Critical Challenges in Adopting DevOps Culture in Software Development: A PLS-SEM Perspective}. IEEE Access  \textbf{11} (2023)

\bibitem{KoziolekKR11}
Koziolek, A., Koziolek, H., Reussner, R.H.: {PerOpteryx: automated application of tactics in multi-objective software architecture optimization}. In: 7th International Conference on the Quality of Software Architectures. pp. 33--42. {ACM} (2011)

\bibitem{MePL17}
Me, G., Procaccianti, G., Lago, P.: {Challenges on the Relationship between Architectural Patterns and Quality Attributes}. In: {IEEE} International Conference on Software Architecture. pp. 141--144. {IEEE} Computer Society (2017)

\bibitem{MeijerTA24}
Meijer, W., Trubiani, C., Aleti, A.: {Experimental evaluation of architectural software performance design patterns in microservices}. J. Syst. Softw.  \textbf{218} (2024)

\bibitem{Razzaq20}
Razzaq, A.: {A Systematic Review on Software Architectures for IoT Systems and Future Direction to the Adoption of Microservices Architecture}. {SN} Comput. Sci.  \textbf{1}(6), ~350 (2020)

\bibitem{DSS1}
Siamak~Farshidi, Slinger~Jansen, R.d.J., Brinkkemper, S.: {A decision support system for software technology selection}. Journal of Decision Systems  \textbf{27}(sup1),  98--110 (2018)

\end{thebibliography}
\end{document}